\documentstyle[twocolumn,aps]{revtex}

\input{epsf}
\begin{document}

\draft
\title{Dynamic depletion in a Bose condensate via a sudden increase
of the scattering length}
\author{C. K. Law, P. T. Leung and M.-C. Chu}
\address{{Department of Physics,
The Chinese University of Hong Kong,}\\
{Shatin, NT, Hong Kong, China}}
\date{\today}
\maketitle

\begin{abstract}
We examine the time-dependent quantum depletion of a trapped Bose
condensate arising from a rapid increase of the scattering length.
Our solution indicates that a significant buildup of incoherent
atoms can occur within a characteristic time short compared with
the harmonic trap period. We discuss how the depletion density and
the characteristic time depend on the physical parameters of the
condensate.

\end{abstract}
\vspace{10mm} 

Bose-Einstein condensates (BEC) of atomic gases with tunable
self-interaction strengths can be realized by applying strong
magnetic fields near a Feshbach resonance
\cite{ketterle1,cornish}. Recent experiments have demonstrated
dramatic dynamic features by manipulating the values of s-wave
scattering lengths. For examples, a change of the interaction sign
from repulsive to attractive can trigger a condensate collapse
\cite{roberts,donley}, a sweep of the magnetic fields across a
resonance leads to an enhanced rate of inelastic collisions
\cite{ketterle2}. More recently Claussen {\it et al.} employed a
magnetic field pulse to turn on a large scattering length in an
extreme short time \cite{xxx}. Their observation of condensate
loss and its counterintuitive dependence on the pulse length
suggest rich microscopic dynamics not captured by the mean field
theory.

In this paper we address the interesting question how a sudden
increase of the self-interaction strength affects the macroscopic
quantum coherence. In particular we will examine the dynamic
creation of incoherent atoms from the condensate (i.e., a
time-dependent depletion effect) based on a two-body interaction
model. Unlike the study of stationary systems \cite{java}, the
calculation of depletion in nonstationary systems is a more
difficult task because one needs to follow the time evolution of
all collective excitation modes (assuming these modes are well
defined in the remote past when the system is stationary)
\cite{castin}. In this paper we present an approximate solution to
tackle a time-varying scattering length problem. We find that the
incoherent atom density can quickly build up if the change of
scattering length is sufficiently large and rapid. Therefore a
sudden change of interaction strength could lead to observable
decoherence effects in the condensate. Our solution indicates the
time scale of the depletion dynamics and the conditions that
determine when the mean field theory breaks down.

To begin we consider a Bose-Einstein condensate of $N$ atoms
confined in a spherical harmonic potential with a trap frequency
$\omega_T$. The time-dependence of the s-wave scattering length
$a(t)$ is controlled by an external magnetic field pulse. We
assume that the magnetic field does not cross any resonance, and
possible molecule formation processes will not be considered.
Under the diluteness assumption of the system $n a(t)^3 \ll 1$
($n$ is the atom number density), the interaction between atoms is
modelled by a two-body $\delta-$potential. The second quantized
time-dependent Hamiltonian of our model is given by,
\begin{equation}
H = \int {d^3 x} \left( {\hat \Psi ^\dag  h_0 \hat \Psi  +
\frac{{2\pi \hbar ^2 a(t)}}{m} \hat \Psi ^\dag  \hat \Psi ^\dag
\hat \Psi \hat \Psi } \right)
\end{equation}
Here $\hat \Psi$ is the atomic field operator, $m$ is the atomic
mass and $h_0 \equiv { - \frac{{\hbar ^2 }} {{2m}}\nabla ^2  +
\frac{1}{2}m\omega_T ^2 r^2 }$ corresponds to the trap Hamiltonian
for a single particle. We note that the validity of replacing the
real interatomic potential by the $\delta-$potential in time
dependent situations is a subtle issue. However, it is plausible
that the $\delta-$potential is valid if the rate of change of the
system is small compare with a characteristic rate $\tau_s^{-1}$
that atoms need to adjust their relative wave functions
\cite{leggett}. The $\tau_s$ is typically a short time depending
on the scattering length, we will discuss this issue later in this
paper.

In this paper we assume that the condensate is stationary
initially with a zero scattering length, i.e., $a(0)=0$. Such an
initial condition can be achieved by turning off the
self-interaction of the condensate adiabatically via a magnetic
field \cite{xxx}. At later time, $a(t)$ is quickly raised to a
peak value $\bar a >0 $. The rise time interval is sufficiently
short (but longer than $\tau_s$) such that the evolution of the
condensate is negligible during the turn on. This allows us to
approximate $a(t)$ in the form
\begin{equation}
a(t) \approx \bar a \theta(t),
\end{equation}
where $\theta$ is the step function.

We decompose the field operator into a dominant coherent part and
a small fluctuation part, i.e., $\hat \Psi = \sqrt N \Phi_0 + \hat
\psi.$ We require the condensate wave function $\Phi_0$ to obey
the time dependent mean field equation,
\begin{equation}
i\hbar \frac{{\partial \Phi_0 }}{{\partial t}} = h_0 \Phi_0  +
g(t) \left| \Phi_0 \right|^2 \Phi_0,
\end{equation}
where $g(t) \equiv 4\pi \hbar ^2 a(t)N/ m$. Then according to the
Heisenberg equation of $\hat \Psi$, the fluctuation $\hat \psi$
obeys the equation,
\begin{equation}
i\hbar \frac{{\partial \hat \psi }}{{\partial t}} = \left( {h_0  +
2g(t) \left| \Phi_0  \right|^2 } \right)\hat \psi  + g(t) \Phi_0
^2 \hat \psi ^\dag.
\end{equation}
Note that in writing Eq. (4) we have assumed that $\hat \psi$ is
small so that second and higher order terms of $\hat \psi$ are
discarded.

Our task is to solve the time evolution of the operator $\hat
\psi$ and examine the growth of incoherent atom density
$\left\langle {\hat \psi ^{\dag}(x,t)\hat \psi (x,t)}
\right\rangle $. First we need to specify the initial conditions
for $\Phi_0$ and $\hat \psi$.
Since the system is noninteracting until $t=0$, the initial
condensate wave function $\Phi_0$ is just the ground state of the
harmonic trap. The corresponding quantum fluctuations at zero
temperature is described by a vacuum state  $\left| {vac}
\right\rangle$ associated with $\hat \psi (\hat x,0)$, i.e., $\hat
\psi (\hat x,0)\left| {vac} \right\rangle  = 0$. Using a standard
expansion in terms of the single particle eigenfunctions of the
trap, the initial conditions are given by,
\begin{eqnarray}
&& \Phi_0(\vec x,0)=f_0(\vec x) \\
&& \hat \psi (\vec x, 0) = \sum\limits_n
{\hat b_n  f _n } (\vec x).
\end{eqnarray}
Here $\hat b_n$ is the annihilation operator associated with the
trap eigenfunction $f_n$ obtained from
$h_0 f_n =\hbar \omega_n f_n$,
with $f_0$ being the lowest state. The initial state $\left| {vac}
\right\rangle$ is the vacuum of $\hat b_n$.

The linear structure of Eq. (4) suggests a general solution in the
form
\begin{equation}
\hat \psi (\vec x,t) = \sum\limits_n^{} {\hat b_n } u_n (\vec x,t)
+ \hat b_n^\dag  v_n^* (\vec x,t),
\end{equation}
where the $\hat b_n$ and $\hat b_n^{\dag}$ are time-independent
operators defined in Eq. (6). A direct substitution
of (7) in (4) gives the equation of motion of the mode functions
$u_n (\vec x,t)$ and $v_n (\vec x,t)$:
\begin{eqnarray}
i\hbar \frac{{\partial u_n }}{{\partial t}} &=& \left( { h_0  +
2g(t)\left| {\Phi _0 (\vec x,t)} \right|^2 } \right)u_n \nonumber
\\
&& + g(t)\Phi _0^2 (\vec x,t)v_n \\ i\hbar \frac{{\partial v_n
}}{{\partial t}} &=&  - \left( { h_0 + 2g(t)\left| {\Phi _0 (\vec
x,t)} \right|^2 } \right)v_n \nonumber
\\ && -g(t)\Phi _0^{*2} (\vec x,t)u_n.
\end{eqnarray}
To match the initial condition (6), we set
\begin{eqnarray}
&& u_n (\vec x,0) = f_n (\vec x)
\\
&& v_n (\vec x,0) = 0.
\end{eqnarray}
Thus the operator equation (4) with the initial condition (6) can
now be replaced by the $c-$number equations (8-11). For later
purpose, it is convenient to introduce the quantity $\lambda (\vec
x)\equiv 4\pi \hbar \bar a Nf _0^2 (\vec x) /m$ that characterizes
the position dependent self-interaction in frequency unit.

In this paper we present a solution in the short time domain $0\le
t \le T_0 $ in which $T_0$ is defined by $T_0 \equiv 1/ \lambda
(0)$ with $\omega_T T_0 \ll 1$. In this domain the condensate wave
function does not have enough time to evolve spatially. The
condensate can only pick up a self-modulated phase arising from
the self-interaction, i.e.,
\begin{equation}
\Phi _0 (\vec x,t) \approx f_0 (\vec x)e^{ - i\lambda (\vec x)t}.
\end{equation}
Note that the trival phase factor $e^{-i\omega_T t/2}$ due to the
trap ground state energy is ignored since $\omega_T t \ll 1$ under
the stated conditions above. We can easily verify the solution by
a substitution in Eq. (3). The approximation we made is the
discard of terms involving the spatial gradient of $e^{ - i\lambda
(\vec x)t}$. These terms would describe the expansion of the
condensate but they only become significant in a longer time scale
$ \approx 1/ \sqrt{\omega_T \lambda(0)} > T_0$, which is beyond
our specified time domain. We have also performed numerical tests
that show Eq. (12) has good agreement with the exact numerical
solution of Eq. (3) within the specified time domain.

We can solve Eq. (8) and (9) in a similar fashion. We write
\begin{eqnarray}
&& u_n (\vec x,t) = f_n (\vec x)U_n (\vec x,t) + \delta _{nu}
(\vec x,t) \\
&& v_n (\vec x,t) = f_n (\vec x)V_n (\vec x,t) +
\delta _{nv} (\vec x,t)
\end{eqnarray}
where the functions $U_n$ and $V_n$ are defined by,
\begin{eqnarray}
i\hbar \frac{{\partial U_n }}{{\partial t}} &=& \left( {\hbar
\omega _n  + 2g(t)\left| {\Phi _0 (\vec x,t)} \right|^2 }
\right)U_n \nonumber \\ && + g(t)\Phi _0^2 (\vec x,t)V_n
\\
i\hbar \frac{{\partial V_n }}{{\partial t}} &=&  - \left( {\hbar
\omega _n  + 2g(t)\left| {\Phi _0 (\vec x,t)} \right|^2 }
\right)V_n \nonumber \\ && - g(t)\Phi _0^{*2} (\vec x,t)U_n
\end{eqnarray}
subjected to the condition $U_n(\vec x,0)=1$ and $V_n(\vec x,0) =
0$. We shall show that the $\delta _{nu}$ and $\delta_{nv}$ are
small correction terms in the short time domain that we are
concerned.

Let us write down the explicit form of $U_n$ and $V_n$. Since Eq.
(15) and (16) do not involve any gradient operator, different
spatial coordinates are decoupled. At each position the equations
behave as a two-level system that can be exactly solved:
\begin{eqnarray}
&& U_n (\vec x,t) = e^{ - i\lambda (\vec x) t} \left[ {\cos \Omega
_n (\vec x)t - i\frac{{\omega _n  + \lambda (\vec x)}}{{\Omega _n
(\vec x)}}\sin \Omega _n (\vec x)t} \right] \nonumber \\
\\
&& V_n (\vec x,t) = ie^{i\lambda (\vec x) t}
\frac{{\lambda (\vec x)}}
{{\Omega _n (\vec x)}}\sin \Omega _n (\vec x)t
\end{eqnarray}
which satisfy the normalization $|U_n|^2-|V_n|^2=1$, and the
quantity
\begin{equation}
\Omega _n (\vec x) = \sqrt {\omega _n \left( {\omega _n  + 2
\lambda (\vec x)} \right)}
\end{equation}
can be interpreted as a form of frequency analogous to the Rabi
oscillations in two-level systems.

To show $\delta_{nu}$ and $\delta_{nv}$ are small, we note that
they are both zero initially. Their equations of motion can be
obtained by a direct substitution of Eq. (13) and (14) into (8)
and (9). This leads to a set of linear equations similar to Eq.
(8) and (9) but with  inhomogeneous terms as
\begin{eqnarray}
&& i\frac{{\partial \delta _{nu} (\vec x,t)}}{{\partial t}} =  -
\frac{\hbar }{m}\left( {\nabla U_n  \cdot \nabla f_n  -
\frac{1}{2}f_n \nabla ^2 U_n } \right) + ...
\\
&& i\frac{{\partial \delta _{nv} (\vec x,t)}}{{\partial t}} =
\frac{\hbar }{m}\left( {\nabla V_n  \cdot \nabla f_n  -
\frac{1}{2}f_n \nabla ^2 V_n } \right) + ...
\end{eqnarray}
These inhomogeneous terms involve the spatial gradient of $f_n$,
$U_n$ and $V_n$. Since $U_n$ and $V_n$ are smooth functions in the
length scale of the trap ground state, the dominant contribution
comes from ${\nabla f_n }$ for high excited modes. We can estimate
the size of $\delta_{nv}$ using the first order approximation
which is a time integration of the inhomogeneous term $\nabla V_n
\cdot \nabla f_n$. Up to the time $t = T_0$, we find that
$\delta_{nv}$ is only a small fraction of $f_nV_n$. The fraction
is typically of the order
$\left( {\omega_T T_0 } \right)^{1/2} \ll 1$
\cite{remarks}. A similar conclusion applies to $\delta_{nu}$.

The functions $f_nU_n$ and $f_nV_n$ are approximate solutions of
$u_n$ and $v_n$. The approximation comes from the fact that
$\nabla ^2 \left( {f_n U_n } \right) \approx U_n \nabla ^2 f_n$
and $\nabla ^2 \left( {f_n V_n } \right) \approx V_n \nabla ^2
f_n$ are the leading contributions in the Laplacians for slowly
varying (spatially) functions $U_n$ and $V_n$. The correction
terms can remain small under the conditions stated in the previous
paragraph. We have tested the approximate scheme beyond the first
order consideration by solving Eq. (8) and (9) numerically in a
one-dimensional model, the exact results are in good agreement
with those obtained by the approximate method.

Now we come to the main result of our paper. The time-development
of incoherent atom density after the change of the scattering
length is given by,
\begin{eqnarray}
\left\langle {\hat \psi ^\dag  \left( {\vec x,t} \right)\hat \psi
\left( {\vec x,t} \right)} \right\rangle  &=& \sum\limits_n^{}
{\left| {v_n \left( {\vec x,t} \right)} \right|^2 } \nonumber \\
&& \approx \sum\limits_n^{} {\left| {f_n \left( {\vec x} \right)}
\right|^2 } \left| {V_n \left( {\vec x,t} \right)} \right|^2.
\end{eqnarray}
With the help of Eq. (18), we have,
\begin{eqnarray}
\left\langle {\hat \psi ^\dag  \left( {\vec x,t} \right)\hat \psi
\left( {\vec x,t} \right)} \right\rangle   \approx
\sum\limits_n^{} {\left| {f_n \left( {\vec x} \right)} \right|^2 }
\frac{{\lambda ^2 \left( {\vec x} \right)}}{{\omega _n \left(
{\omega _n  + 2\lambda \left( {\vec x} \right)} \right)}}
\nonumber \\ \times \sin ^2 \left[ {\sqrt {\omega _n \left(
{\omega _n + 2\lambda \left( {\vec x} \right)} \right)} t}
\right].
\end{eqnarray}
For a spherical harmonic trap, each single particle state $n$ is
described by three quantum numbers ($n_1 ,n_2 ,n_3$) in Cartesian
coordinates, and the corresponding frequency $ \omega_n
=(n_1+n_2+n_3+3/2)\omega_T$. Therefore Eq. (23) is a triple
summation.

Let us calculate the incoherent atom density at the trap center
$\vec x =0$. It is a good approximation (particularly for high
modes) that
\begin{eqnarray}
&& \left| {f_{2j_1 ,2j_2 ,2j_3 } \left( 0 \right)} \right|^2
\approx 0.09 \times l_0^{ - 3} \nonumber \\ && \ \ \ \ \times
\left[ {\left( {2j_1 + \frac{1}{2}} \right)\left( {2j_2  +
\frac{1}{2}} \right)\left( {2j_3  + \frac{1}{2}} \right)}
\right]^{ - 1/2}
\end{eqnarray}
where $l_0 = \sqrt{\hbar /m \omega_T}$  and we need only to
consider even modes at the origin. Eq. (24) is obtained from the
semi-classical construction of the trap eigenfunctions. One can
check that Eq. (24) is fairly accurate even for the ground state.
Next we change the variables $j_i$ by introducing: $ \tilde k_i^2
= (4j_i  + 1)\omega _T T_0/2 $ for $i=1,2,3$. We find that the
function inside the summation (23) changes smoothly with $\tilde
k_i$. Therefore we may replace the discrete sum (23) by an
integral that can further be simplified using the spherical
symmetry. In order to compare the fluctuations with the mean-field
density, we define $\Delta (t) \equiv \left\langle {\hat \psi
^\dag (0,t)\hat \psi (0,t)} \right\rangle /n_0$ which is the ratio
of the central incoherent atom density to the condensate density
$n_0=Nf_0^2(0)$. We find that
\begin{equation}
\Delta (t) \approx  6.3 ({n_0 \bar a^3
})^{1/2}\int\limits_0^\infty {dk\frac{{\sin ^2 ( kt\sqrt {2 + k^2
} /T_0)}}{{2 + k^2 }}}
\end{equation}
where $k^2= \tilde k_1^2+\tilde k_2^2+\tilde k_3^2$.  The integral
in Eq. (25) is an increasing function of time, but it must be less
than $\pi /2\sqrt 2 \approx 1.1$. At the time $t=T_0$ the integral
reaches $0.53$ that gives $\Delta \approx 3.4(n_0 \bar a^3)
^{1/2}$ \cite{remark2}. This result is about twice the value that
one would obtain from a (stationary) homogeneous condensate at
zero temperature \cite{huang}. Note that $\Delta (t)$ depends
explicitly on the gas parameter $n_0\bar a^3$ in Eq. (25). If
$\Delta (t)$ is comparable to one, then both the mean field theory
and the linearized quantum theory break down because quantum
fluctuations are no longer small. For a general position $\vec x$,
we can work out a similar integral. The spatial profile can be
well fitted by a Gaussian:$ \left\langle {\hat \psi ^\dag (\vec
x,t)\hat \psi (\vec x,t)} \right\rangle \approx \left\langle {\hat
\psi ^\dag (0,t)\hat \psi (0,t)} \right\rangle e^{ - 3r^2
/2l_0^2}$.

Finally we would like to discuss the validity of the model (1). A
necessary condition that the usual pseudopotential approach can be
generalized to time-dependent situations is that the time-varying
parameters of the system do not excite high relative wave vectors
between atoms. This condition ensures $ p a/\hbar \ll 1$ (where
$p$ is the relative momentum) required for the s-wave
approximation. There is no simple way to check the condition for
general interatomic potentials, but it is useful to consider the
hard sphere model as an illustrative example. In the relative
coordinate of two hard spheres, the expanding spheres are
described by a moving boundary condition. If the radial velocity
is $\dot a$, a slow incoming wave will pick up a wavevector of
magnitude roughly $m_r \dot a / \hbar$ by reflection, where
$m_r=m/2$ is the reduced mass. Therefore $m_r \dot a a / \hbar \ll
1$ is the required condition that specifies how slow the speed
should be. If we consider $a(t)=\bar a t/t_{rise}$ is linear in
time during a rise period $t_{rise}$, then the condition is
satisfied if $m_r \bar a^2 / \hbar \ll t_{rise}$. Quite
remarkably, the characteristic time $\tau_s \equiv m_r \bar a^2 /
\hbar$ estimated here agrees with the one suggested by Leggett
under a general consideration \cite{leggett}.

To give a physical example, we consider a $^{85}$Rb condensate
with $10^5$ atoms in a spherical trap with $\omega_T = 2 \pi
\times 10$ Hz. If the scattering length $\bar a = 30$ nm, then
$T_0=80 \mu s$ and $\tau_s = 1 \mu s$ is a short time. In this
example, the gas parameter $n_0\bar a^3 = 0.001$ and $\Delta =0.11
$ at $t=T_0$. This means a $11\%$ depletion of condensate density
at the trap center in $80 \mu s$. We may compare this example with
the case with a smaller scattering length $\bar a = 3$ nm (same
particle number and trap frequency) that only gives $0.36\%$
depletion density in $800 \mu s$.

In conclusion we made use of the model (1) to analyze the dynamic
depletion in a Bose condensate after a sudden increase of the
self-interaction strength. The depletion is a quantum decoherence
process originated from the vacuum fluctuations associated with
$\hat \psi$. Our solution (23) and (25) indicate how the depletion
density increases as a function of time. We discover that
$T_0=m/4\pi \hbar \bar a Nf _0^2 ( 0)$ is an important time scale
for the buildup of incoherent atoms for a trapped condensate with
$\omega_T T_0 \ll 1$. It is interesting to rewrite $T_0$ in the
form: $T_0 = \omega_T^{-1} (n_0 \bar a^3) ^{-1/3} (n_0 l_0^3)
^{-2/3}/4 \pi$. This indicates the relationship between the
characteristic time $T_0$ and different parameters, including the
gas parameter. Finally, we point out that the recent experiment at
JILA \cite{xxx} reaching $n_0\bar a^3 \approx 0.3$ is beyond the
scope of the present theory. Such a strong coupling would give
$\tau_s \approx 39 \mu s$ that is {\em longer} than the rise time
of the magnetic pulse, and so the s-wave approximation model is
questionable. The microscopic processes in such an extreme
situation remain an interesting open topic for investigations.

\acknowledgments We acknowledge the support from an RGC Earnmarked
Grant CUHK4237/01P and the Chinese University of Hong Kong Direct
Grant (Grant Nos: 2060148 and 2060150).


\begin{references}
\bibitem{remarks} Our approximation works even better at $r=0$,
since $\nabla V_n \cdot \nabla f_n$  and $\nabla U_n \cdot \nabla
f_n $ both vanish at the center. The non-zero inhomogeneous terms
$f_n \nabla ^2 V_n$ and $f_n \nabla ^2 U_n$ contribute a
correction of order $\omega_T T_0 $ instead of $\left({\omega_T
T_0} \right)^{1/2}$.
\bibitem{remark2} The growth of $\Delta$ is most significant
within the time $T_0$ after the interaction is turned on.
Extending the duration does not yield a much increase of $\Delta$.
For example, for the same condensate, twice the duration (i.e.,
$2T_0$) gives $\Delta (2T_0) \approx 3.5 (n_0 a^3)^{1/2}$ which is
only about a $3\%$ increase compared with the value at $t=T_0$.
\end{references}
\end{document}